\documentclass[review,number]{elsarticle}
 \usepackage{lineno}

\usepackage{amssymb}
\usepackage{amsmath}
\usepackage{hyperref}
\usepackage{graphicx}
\usepackage{float}

\begin{document}

\begin{frontmatter}

\title{Ultrafast laser pulse heating of metallic photocathodes and its contribution to intrinsic emittance}

\author{J. Maxson, P. Musumeci}
\address{Department of Physics and Astronomy, UCLA, Los Angeles, California 90095, USA}
\author{L. Cultrera}
\address{CLASSE, Cornell University, Ithaca, NY 14850}

\author{S. Karkare, H. Padmore}
\address{Lawrence Berkeley National Laboratory, Berkeley, CA 94720}

\begin{abstract}
The heating of the electronic distribution of a copper photocathode due to an intense drive laser pulse is calculated under the two-temperature model using fluences and pulse lengths typical in RF photoinjector operation. Using the finite temperature-extended relations for the photocathode intrinsic emittance and quantum efficiency, the time-dependent emittance growth due to the same photoemission laser pulse is calculated. This laser heating is seen to limit the intrinsic emittance achievable for photoinjectors using short laser pulses and low quantum efficiency metal photocathodes. A pump-probe photocathode experiment in a standard 1.6 cell S-band gun is proposed, in which simulations show the time dependent thermal emittance modulation within the bunch from laser heating can persist for meters downstream and, in principle, be measured using a slice emittance diagnostic.

\end{abstract}


\end{frontmatter}

\section{Introduction}\label{intro}

Photocathode electron sources are the electron source of choice for a wide variety of high brightness electron beam applications, including (but not limited to) free electron lasers (FELs) \cite{EmmaAkreArthurEtAl2010}, energy recovery linacs \cite{NeilBohnBensonEtAl2000, GullifordBartnikBazarovEtAl2015}, as well as ultrafast electron microscopy and diffraction(UEM/D) \cite{MusumeciMoodyScobyEtAl2010, LiMusumeci2014}. The advent of emittance compensation \cite{SerafiniRosenzweig1997}, utilizing detailed knowledge of the initial spatial laser profile and space-charge dominated beam dynamics, has allowed the production of photoemitted beams with significant charge density and brightness dominated  by the photocathode emittance, defined as:

\begin{equation}
\epsilon_{x, i} = \sigma_x \sqrt{ \frac{\text{MTE}}{m c^2} },
\end{equation}

\noindent where $\sigma_x$ is the rms spot size of the laser on the photocathode along a transverse cartesian coordinate $x$, $mc^2$ is the electron rest energy, and where $\text{MTE}$ is the mean transverse energy of photoemission, analogous to the temperature of the photoemitted electrons. Generally, the initial laser size is determined by the beam's space charge dynamics, and the MTE is the single material parameter that encompasses the momentum spread induced in the electron's photoexcitation, transport to the surface (including any scattering mechanisms), and escape into the vacuum (including any contribution from surface nonuniformity). The prediction of the photocathode quantum efficiency and MTE requires a detailed knowledge of each of the above processes, and is an active area of research for both metallic and semiconducting emitters. 

Metallic photocathodes are used widely in high brightness photoinjectors due to their insensitivity to poor vacuum conditions and their prompt temporal response.  The prompt response times  ($<$ 50 fs) of metallic photocathodes enables novel beam dynamics methods, such as the use of the so called ``blow-out" emission regime \cite{MusumeciMoodyEnglandEtAl2008, LuitenEtAl2004}, in which an ultrafast laser pulse creates an initial charge distribution which is longitudinally thin and radially wide, which expands under space charge forces to create a uniformly filled ellipsoid downstream. The blowout regime can be driven via linear or nonlinear \cite{MusumeciCultreraFerrarioEtAl2010} photoemission processes. 

However, the low quantum efficiency of metals requires high laser fluences, often on the order of the damage/ablation fluence of the metallic photocathode surface (10s of mJ/cm$^2$), coupled with the ultrafast ($\sim 100 $ fs) pulses yields very large laser intensities (10s of GW/cm$^2$). At such intensities, it was proposed \cite{AnisimovKapeliovichPerelman1974} and measured via time resolved reflectivity/transmissivity \cite{Elsayed-AliNorrisPessotEtAl1987}  or two photon photoemission spectroscopy \cite{Fann1992}  that on the sub-ps time scale of the laser pulse the temperature of the electronic distribution  is effectively isolated from the lattice, and can increase to several thousand Kelvin, due in part to the large difference between the electron and lattice heat capacities.  After this initial rise, equilibrium between the electrons and lattice is reached on the ps-timescale via electron-phonon scattering.  In this work, we seek to calculate the extent to which a single ultrafast photoemission laser pulse's heating of the photocathode electronic distribution increases the intrinsic emittance of the electrons it emits.

To understand the conceptual role of the electronic temperature on the photoemission MTE, we first review previous analytic calculations of metallic photocathode properites. The MTE and quantum efficiency of metal photocathodes was given by the well-known calculation by Dowell and Schmerge \cite{DowellSchmerge2009} using a free-electron Fermi gas model with flat density of states and at zero temperature. For photon energy well above the workfunction $h\nu\gg \phi$, it was shown that $\text{MTE} = \left( h\nu - \phi\right)/3$.  This calculation was extended to include the effects of finite temperature and a realistic density of states in \cite{VecchioneDowellWanEtAl2013, FengNasiatkaWanEtAl2015}, in which it was shown that for photon energies at or below work function, $\text{MTE} \approx k T_e $, or the temperature of the electronic distribution. Hence, it is clear that if large enough, the electronic distribution temperature may contribute significantly to the emitted electron's MTE. Near threshold, the MTE is dominated by the electronic temperature, and hence any heating of the electron distribution via laser illumination directly limits the lowest achievable MTE. This ultrafast laser heating effect, which for a given cathode material depends primarily on the laser fluence, has not been taken into account in models of photoemission brightness before, and would manifest itself as a time dependent intrinsic emittance in the bunch.    

This work will proceed as follows. First, we will calculate the rise in electronic distribution temperature from the well known two temperature model (TTM), using parameters for a copper photocathode. Then, we will calculate the time dependent change in electron beam emittance off the cathode using the extended Dowell-Schmerge relations, and then demonstrate how this effect may limit the minimum achievable intrinsic emittance in photoinjectors. Finally, we will propose a pump-probe experiment to measure this effect, and show results from space charge simulations with a photoemitted beam that has an intrinsic emittance as a function of time. These simulations indicate that the temperature modulation will persist meters downstream of a standard 1.6 cell normal conducting rf gun, where a slice emittance diagnostic may be employed to measure the temperature modulation of the beam.  

\section{1D two-temperature model}
The TTM, originally proposed in \cite{AnisimovKapeliovichPerelman1974}, treats the electrons and lattice as separate thermal subsystems that interact via an electron-phonon scattering term. In one spatial dimension, the TTM  is given by:

\begin{align}
C_e(T_e) \frac{\partial }{\partial t} T_e =& \frac{\partial}{\partial z}\left(K_e(T_e) \frac{\partial}{\partial z} T_e \right) -g(T_e-T_l)+S(t, z)
\label{TTMeq1}
 \\
C_l(T_l) \frac{\partial}{\partial t} T_l =& g(T_e-T_l)
\label{TTMeq}
\end{align} 

where here $C_{e/l}(T_{e/l})$ is the electron/lattice heat capacity per unit volume, which is a function of the electron/lattice temperature. Both temperatures are functions of time ($t$) and longitudinal position ($z$) into the sample, but we have suppressed this dependence for simplicity. $K_e$ is the electronic thermal conductivity (a function of the electronic temperature), $g$ is the electron-phonon coupling constant, and $S(t, z)$ is the laser intensity source term. For the remainder of the work, we will restrict our discussion to a bulk copper photocathode, a material that has been the subject of a number of ultrafast photoexcitation and TTM modeling studies \cite{Elsayed-AliNorrisPessotEtAl1987, Eesley1986,HohlfeldWellershoffGueddeEtAl2000}.

 We make the standard replacement that the electronic heat capacity varies linearly with the electronic temperature $C_{e} = \gamma T_e$, where $\gamma = 96.6$ $\frac{\text{J}}{\text{m}^3 \text{K}^2}$ \cite{Eesley1986}. For $C_l(T_e)$, the Debye model is applied, where we use the Debye temperature for Cu, $\Theta_D = 343$ K \cite{HohlfeldWellershoffGueddeEtAl2000}. The electron thermal conductivity is also known to depend linearly on the electron temperature, such that we may write $K_e = K_{e0} T_e/T_l$ where $K_{e0} = 401$ $\text{W}/\text{m}\cdot \text{K}$ is the thermal conductivity at 300 K \cite{CorkumBrunelShermanEtAl1988, HohlfeldWellershoffGueddeEtAl2000,WellershoffHohlfeldGueddeEtAl1999}. 
 
 For bulk copper with surface at z=0, we assume a Gaussian temporal distribution for the laser intensity: 
 \begin{equation}
 S(t, z) = \frac{\left(1 -R \right)F_0}{\sqrt{2\pi} \sigma_t d_{p} } \exp\left[ -\frac{\left(t-t_0\right)^2}{2 \sigma_t^2} - \frac{z}{d_{p}}\right]
 \end{equation}
 
 where here $R$ is the reflection coefficient, $F_0$ is the incident laser fluence,  $\sigma_t$ the laser pulse width with $t_0 = 4\sigma_t $, and $d_p$ is the effective penetration depth of the laser energy. It has been previously determined that the effective penetration depth needs to be increased beyond the optical skin depth to include the electron ballistic range, thereby increasing the effective laser-material interaction volume and decreasing the predicted peak temperature \cite{ChenLathamBeraun2002, HohlfeldWellershoffGueddeEtAl2000, WellershoffHohlfeldGueddeEtAl1999}. For gold, the measured values of the electron ballistic range agree reasonably well with the ballistic range calculated theoretically \cite{HohlfeldWellershoffGueddeEtAl2000, AshcroftMermin1976}; for copper, the ballistic range is 70 nm, and the optical skin depth of 800 nm (266 nm) light is 13 nm (12 nm). Hence, throughout this work, we use a constant $d_p = 83$ nm.
 
 Note that the TTM implicitly assumes the existence of an electronic temperature, an assumption that has been shown to be false for low intensity illumination \cite{Fann1992, MuellerRethfeld2013}. In this case, the electron-electron scattering timescale (which produces electronic thermalization), overlaps with the electron-phonon timescale, for which the TTM is no longer applicable.  However, it has been shown that the thermalization time of the electronic distribution decreases with increasing laser fluence: within Fermi-liquid theory, the inverse lifetime of a single hot electron due to electron-electron collisions scales as $\tau_{e-e}^{-1}\sim (kT_e)^2$ \cite{MuellerRethfeld2013}. This is however not a multiple-particle quantity representative of the entire electronic distribution. Alternatively, in \cite{MuellerRethfeld2013}, a Boltzmann collision integral simulation method was applied using a realistic density of states. The thermalization times in gold for the excitation fluences/temperatures less than or equal to those considered in this work were calculated to range from 100s of fs down to 10s of fs, decreasing with increasing fluence. Thus, for the weak excitation case considered in this work ($\sim 10$ mJ/cm$^2$ and 3 ps rms) the electron temperature may not be well defined at sub-ps timescales \cite{Fann1992}, given the overlap of the electron thermalization timescale with electron-phonon relaxation. Nonetheless, in \cite{Fann1992}, the TTM provides agreement with data after roughly 1 ps. Thus, for the purposes of this work, the assumption of immediate thermalization is reasonable.

 Before solving the TTM numerically for a specific case, we can extract a few analytic scaling limits. To estimate the peak electronic temperature, assuming the laser pulse is short compared to the electron-phonon relaxation time,  we can neglect the contribution of the electronic diffusion and coupling to the lattice ($K_e \rightarrow 0$ and $g \rightarrow 0$). It can then be easily shown that the maximum electronic temperature reached in the sample is:
 
 \begin{equation}
 T_{max} \approx \sqrt{ \frac{2\left(1-R\right) F_0}{\gamma d_p} + T_e^2(t= 0)}
 \label{Tmax}
 \end{equation}
 Note that in this limit, the maximum temperature does not depend on the laser pulse length, but only the absorbed fluence. To estimate the timescale for electron-phonon relaxation, we continue to neglect electron heat conduction, but now consider nonzero $g$. Here we may assume $C_l \rightarrow \infty$ and $T_l =0$, so that that $ dT_e/dt = -g/\gamma$, or a linear temperature decrease with a timescale of $\tau_{e-p} \approx T_{max} \gamma /g$. For copper material parameters ($g = 10^{17}$ W/m$^3 \cdot$ K \cite{Eesley1986}),   irradiation with 10 $\text{mJ}/\text{cm}^2$ absorbed fluence yields a $T_{max} = 5000$ K, with a corresponding $\tau_{e-p}$ of roughly 5 ps.
 
 The inclusion of electron heat conduction decreases the peak electron temperature, and hence also increases the rate of equilibration with the lattice. To estimate the strength of this effect, we first ignore lattice and under the assumption of a point laser source term, the diffusion equation Green's function  yields an effective diffusion length as a function of time of \cite{ WellershoffHohlfeldGueddeEtAl1999}:
 \begin{equation}
 L_d(t) = \sqrt{\frac{2 K_{e0} t}{T_l \gamma}}
 \end{equation}
 
 This length can be viewed as decreasing the source energy density in the material, and after 100 fs this length is $\approx 50 \text{ nm}$. Given that this is comparable to the effective penetration depth, and $T_{max}\propto d_p^{-1/2}$, we expect an order unity correction to the $T_{max}$ and $\tau_{e-p}$. However, note that with heat conduction the peak temperature is now a function of the laser pulse duration, even for pulses much shorter than $\tau_{e-p}$.
 
 \begin{figure}[H]
 	\centering
 	\includegraphics[width=0.7\textwidth]{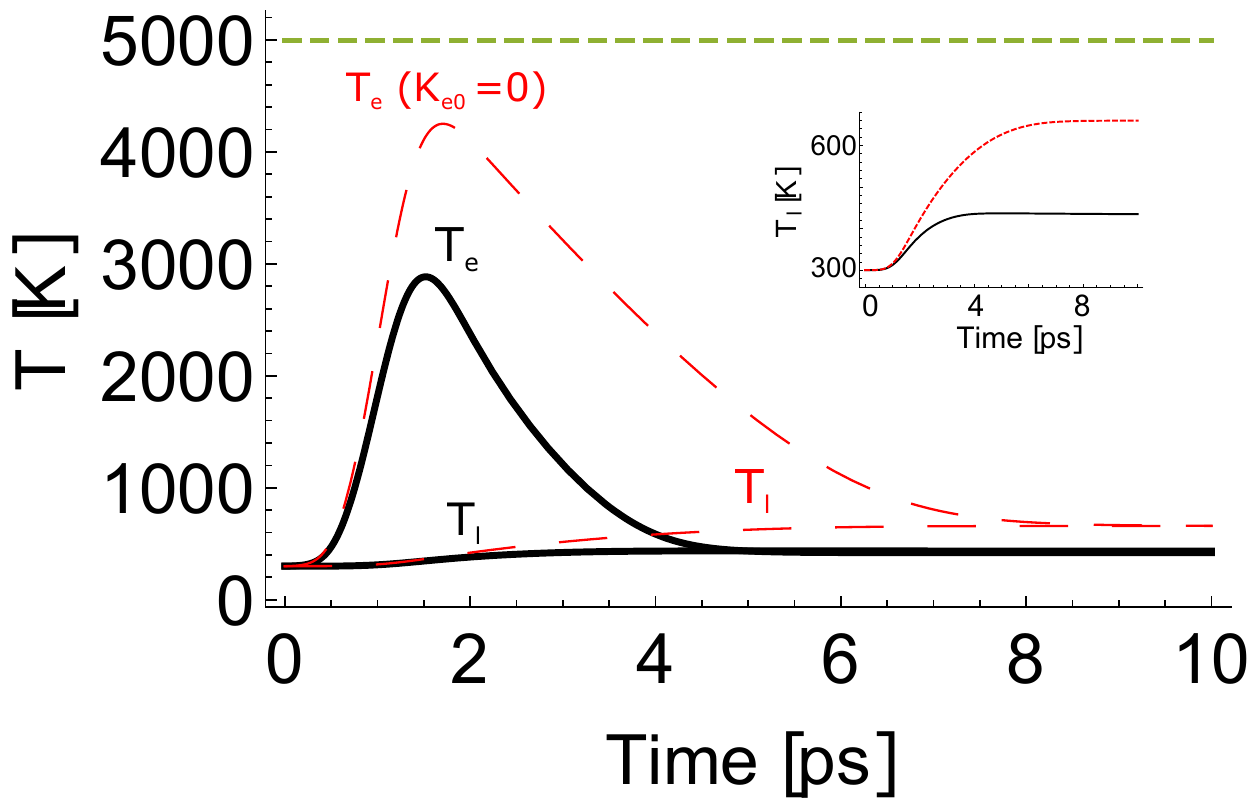}
 	\caption{TTM model surface temperature computed with 10 mJ/cm$^2$ absorbed fluence, showing both electron and lattice temperature at the emitting surface. Red, dashed curves do not include the effects of electron thermal conductivity. Black curves are the full solution of Eqns. \ref{TTMeq1}--\ref{TTMeq}. The green dotted line is the estimation of Eq. \ref{Tmax}. Inset: Close-up of lattice temperatures. }\label{TTM}
 \end{figure}
 
 We solve the TTM numerically for an example where a pulse with 10 mJ/cm$^2$ fluence and $\sigma_t = 300$ fs is absorbed on copper. The solution boundary is extended to $z = 800$ nm,  sufficient to approximate bulk copper, at which point the temperature is held fixed at 300 K.  The electron and lattice temperatures as a function of time are plotted in Fig. \ref{TTM}. The red, dashed curves do not include any effects of thermal conductivity ($K_{e0} \rightarrow 0$ ), and the dotted horizontal line is the estimation of Eq. \ref{Tmax}. The linear decrease of the temperature just after the peak in the case of no thermal conductivity is noteworthy, along with the decrease of both the peak temperature and equilibration time when thermal conductivity is included.

\section{Intrinsic emittance growth}

We calculate the intrinsic emittance increase as a function of time for the temperature profile shown in Fig. \ref{TTM}. We apply the extended Dowell-Schmerge relations for finite temperature presented in \cite{VecchioneDowellWanEtAl2013}.  We assume that the temperature pertinent for the emission is that at the z=0 surface.  

\begin{figure}[ht]
	\centering
	\includegraphics[width=0.6\textwidth]{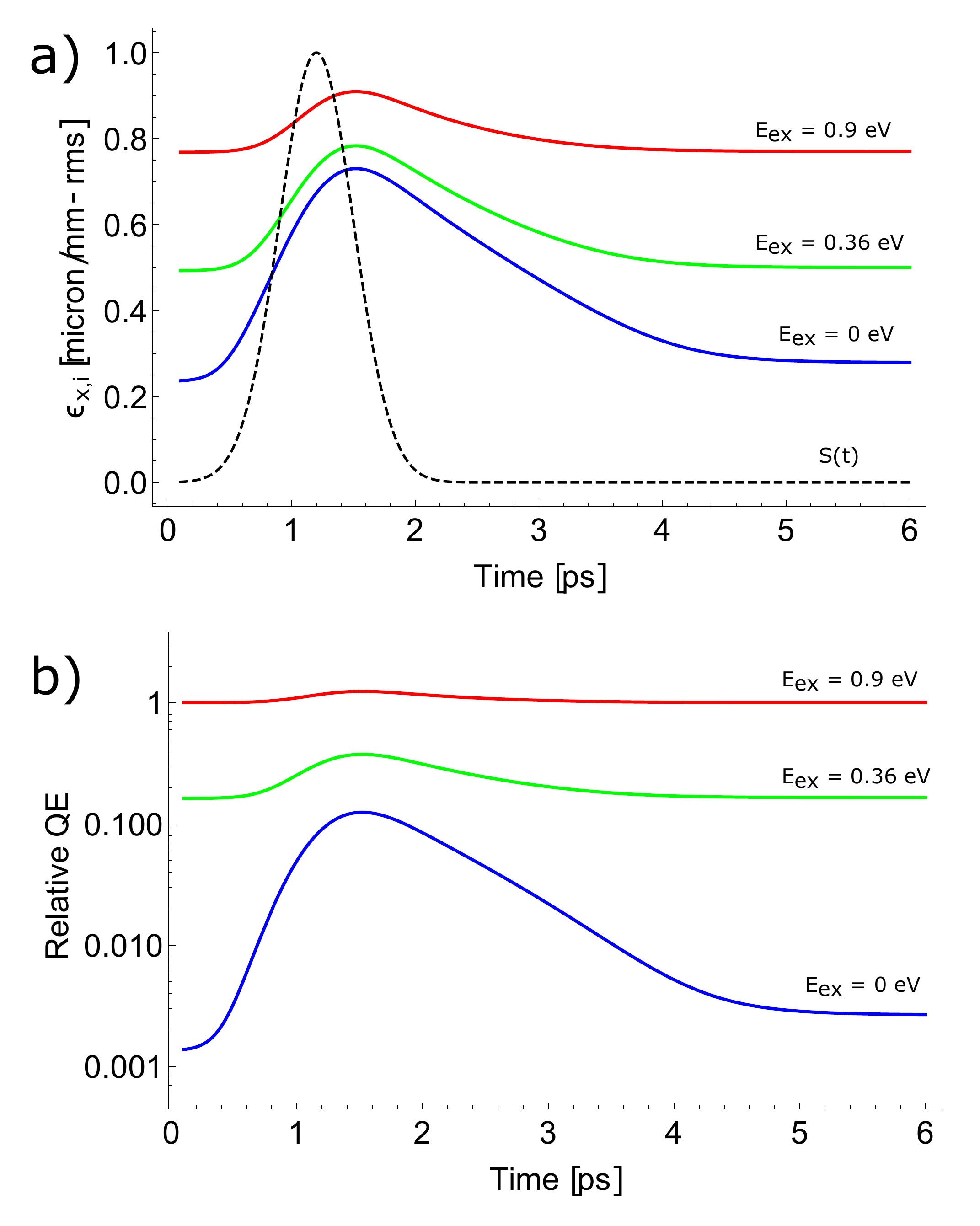}
	\caption{Intrinsic emittance (a) and relative quantum efficiency (b) for $E_{ex} = $ 0.9, 0.36, and 0 eV due to the temperature profile shown in Fig. \ref{TTM} (10 mJ/cm$^2$ absorbed, $\sigma_t = 300$ fs). The relative QE is normalized to the quantum efficiency at $E_{ex} = 0.9$ eV and 300 K. The drive laser intensity profile is shown by the black dotted line.  }\label{etherm}
\end{figure}

The photoemission MTE is given by:

\begin{equation}
\text{MTE} = kT_e \frac{ \text{Li}_3 \left(-\exp \left[E_{ex}/kT_e\right] \right)  }{\text{Li}_2 \left(-\exp \left[E_{ex}/kT_e\right] \right)  }
\label{theo}
\end{equation}
where the excess energy $E_{ex}$ is defined as $h\nu-\phi_{eff}$,  $\phi_{eff}$ is the effective work function including the Schottky lowering due to the applied field, and Li$_n$ is the polylogarithm function of order $n$.  For small excess energy, $\text{MTE}\rightarrow kT_e$, and for large excess energy, $\text{MTE}\rightarrow E_{ex}/3$. 

The quantum efficiency as a function of excess energy and temperature is given by:

\begin{equation}
\text{QE} = C \frac{ \text{Li}_2 \left(-\exp \left[E_{ex}/kT_e\right] \right)}{\text{Li}_2 \left(-\exp \left[E_{f}/kT_e\right] \right)}
\label{QEth}
\end{equation}

where $E_f$ is the Fermi energy (7 eV in Cu), and $C$ is a constant. In general, the prefactor of Eq. \ref{QEth}  also depends on wavelength, via the reflectivity and the electron mean free path \cite{DowellSchmerge2009}, but change in the prefactor from these effects are of order unity and are thus small compared to the modulation due to the polylogarithm.   

The material workfunction is also approximately constant with respect to temperature. Though the workfunction is known to vary with temperature primarily because of lattice effects (thermal expansion and atom vibration) \cite{Kiejna1986}, and the slope of the temperature dependence is in general approximately one Boltzmann constant, and hence for lattice temperatures reaching $\sim 500$ K, the workfunction modulation is much smaller than the contribution from the electronic temperature at the peak.

The intrinsic emittance for multiple values of $E_{ex}$ are plotted in Fig. \ref{etherm} for the temperature profile shown in Fig. \ref{TTM}. Note that the peak emittance modulation is from this effect is apparent even for values of $E_{ex}$ that yield thermal emittances comparable to typical values achieved in experiment ($0.8-1$ $\mu$m/mm) with the 3rd harmonic of 800 nm Ti:Sapphire lasers in s-band rf guns \cite{MusumeciMoodyEnglandEtAl2008,MusumeciCultreraFerrarioEtAl2010}. Furthermore, the relative height of modulation increases with decreasing excess energy. Thus, we see that for ultrashort pulse illumination, the intrinsic emittance is inherently limited by the absorbed fluence.

\begin{figure}[H]
	\centering
	\includegraphics[width=0.5\textwidth]{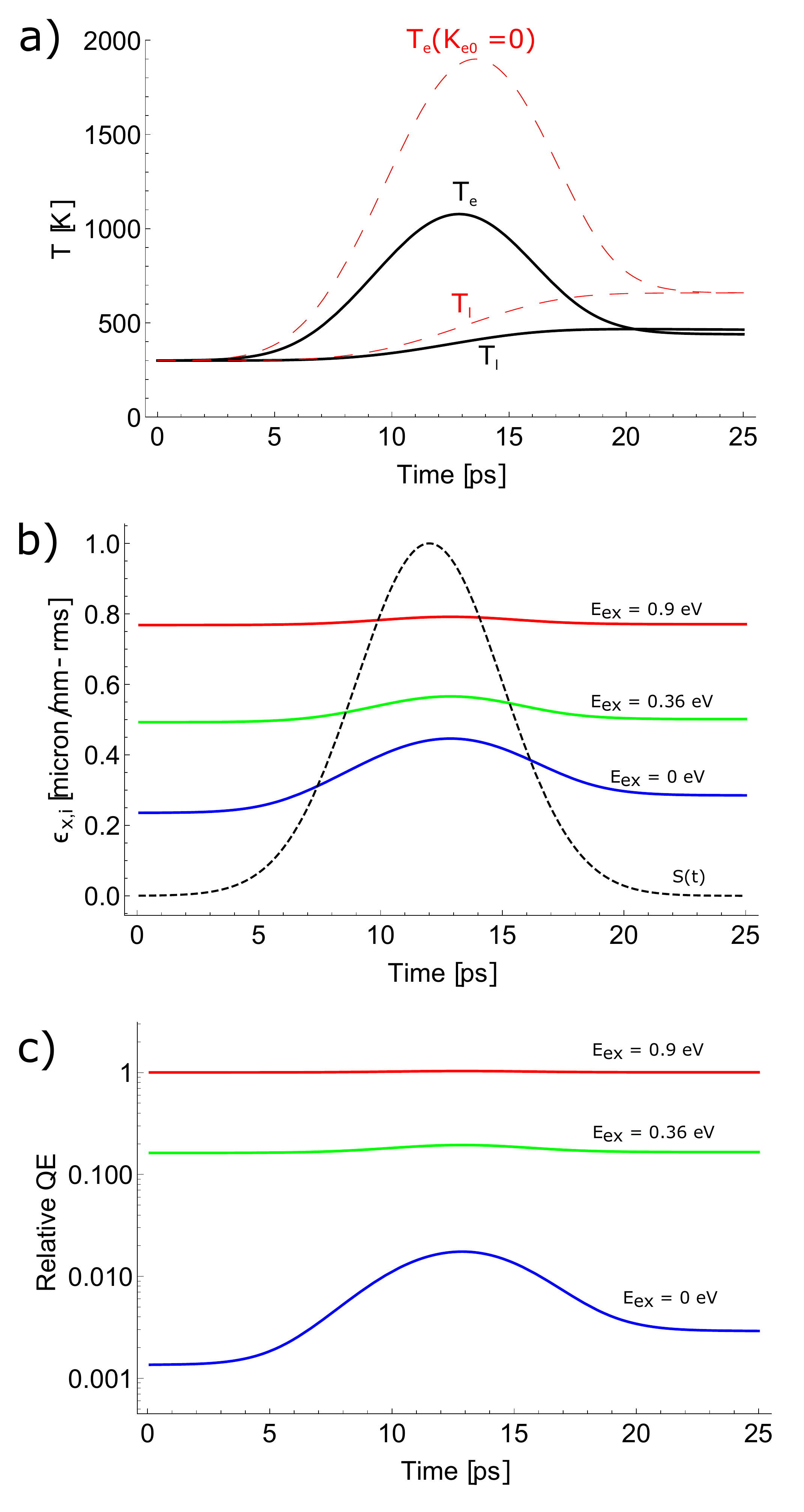}
	\caption{The results of figures \ref{TTM} and \ref{etherm} recomputed for 10 mJ/cm$^2$ (absorbed) and $\sigma_t = 3$ ps, showing  the electron and lattice surface temperature (a) and the corresponding intrinsic emittance (b) and relative QE (c).  The laser intensity profile is shown by the black dotted line.  }\label{ethermps}
\end{figure}

This heating-induced increase of the intrinsic emittance can be reduced by either of two methods. First, utilizing photocathodes of higher quantum efficiency, such as semiconductor photocathodes, directly alleviates the need for high fluence. Such photocathodes, for example GaAs:Ce, Ce$_2$Te, and the alkali antimonides have been shown to have quantum efficiencies in the percent range, greater than the typical QE of copper at 266 nm by roughly 3 orders of magnitude. Secondly, if the application permits, one may utilize longer, multiple ps scale pulses, which reduces the peak temperature significantly via the electron-phonon coupling.

To illustrate the effect of using longer pulses, the case previously considered is recalculated for an order of magnitude longer pulse, $\sigma_t = 3$ ps. The electron and lattice temperatures, as well as the MTE and QE vs time are shown in Fig. \ref{ethermps}. Here, the fluence remains 10 mJ/cm$^2$, and hence the estimate of Eq. \ref{Tmax} remains unchanged. Here, however, the peak temperature only reaches a maximum value of $\sim$1000 K. The MTE and QE well above threshold ($E_{ex} = 0.9$ eV) are nearly unchanged by the temperature rise. However, at threshold ($E_{ex}= 0$), the minimum intrinsic emittance nearly doubles at the time of peak electron temperature.

\begin{figure}[H]
	\centering
	\includegraphics[width=0.7\textwidth]{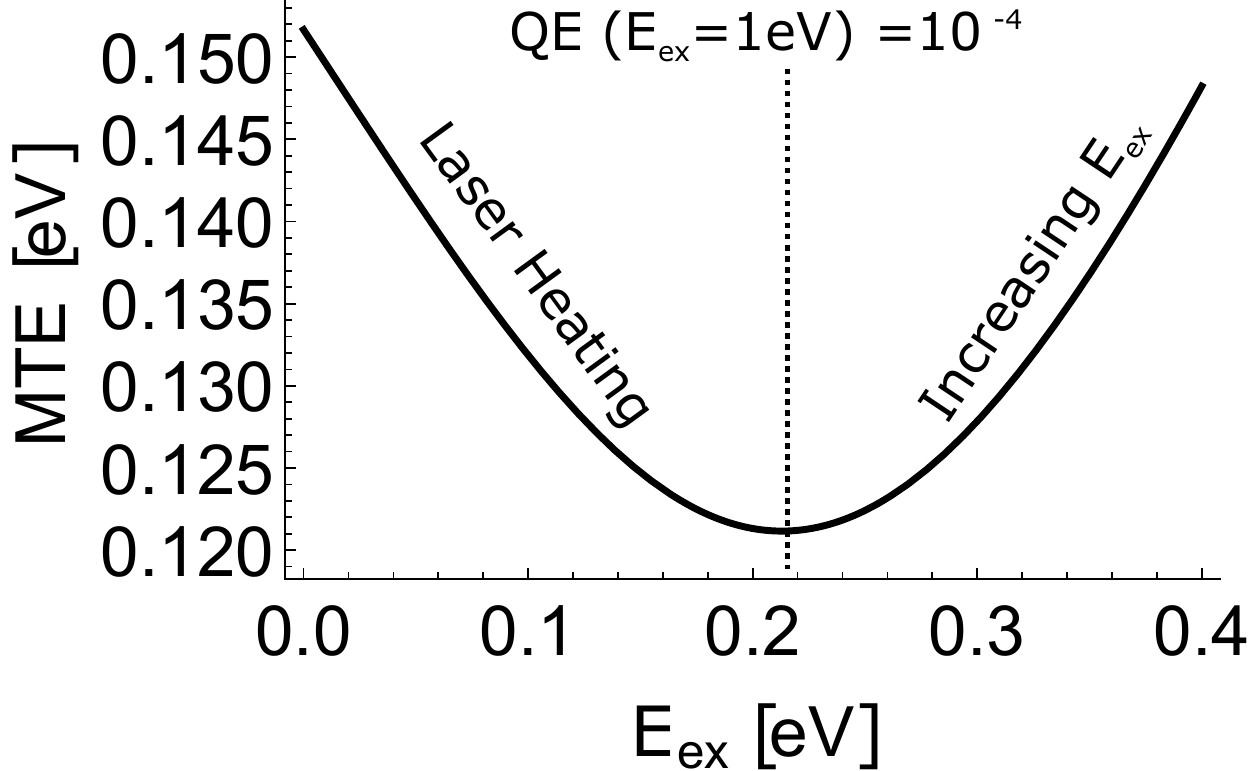}
	\caption{Average MTE as a function of excess energy when extracting 44 pC/mm$^2$ from copper ($\phi_0 = 4.31$ eV) with an extraction field of 50 MV/m and $\sigma_t =  300$ fs. Regions where the MTE is dominated by large $E_{ex}$ and laser heating are shown.    }\label{fluopt}
\end{figure}

The transverse beam charge density is typically set by space charge dynamics to ensure the full charge extraction and emittance preservation. For instance, in the blow-out regime of photoinjector operation, the charge density should be set to $\sigma \leq \alpha E_0 \epsilon_0$, where $E_0$ is the extraction field at the cathode. Here the maximum charge density extractable $E_0 \epsilon_0$ is reduced by a factor $\alpha$ (typically 0.1 for the blowout regime) to make sure the beam emittance is not diluted by image forces during emission \cite{MusumeciMoodyEnglandEtAl2008}. The required absorbed fluence is then:

\begin{equation}
F = \frac{ \alpha E_0 \epsilon_0 \left( E_{ex}+\phi_{eff}\right) \left( 1- R \right) }{ e  \overline{\text{QE}}(F, E_{ex})}
\label{required_fluence}
\end{equation}

\noindent where here we have explicitly written that the average quantum efficiency over the pulse $\overline{\text{QE}} \propto \int \text{QE}(t) S(t, 0) dt$ is a function of the fluence and excess energy. Thus, by numerically tabulating $\overline{\text{QE}}$ given by the TTM and Eq. \ref{QEth}, one may then numerically invert Eq. \ref{required_fluence} for the  relationship between fluence and excess energy that produces the correct charge density.

\begin{figure}[H]
	\centering
	\includegraphics[width=0.7\textwidth]{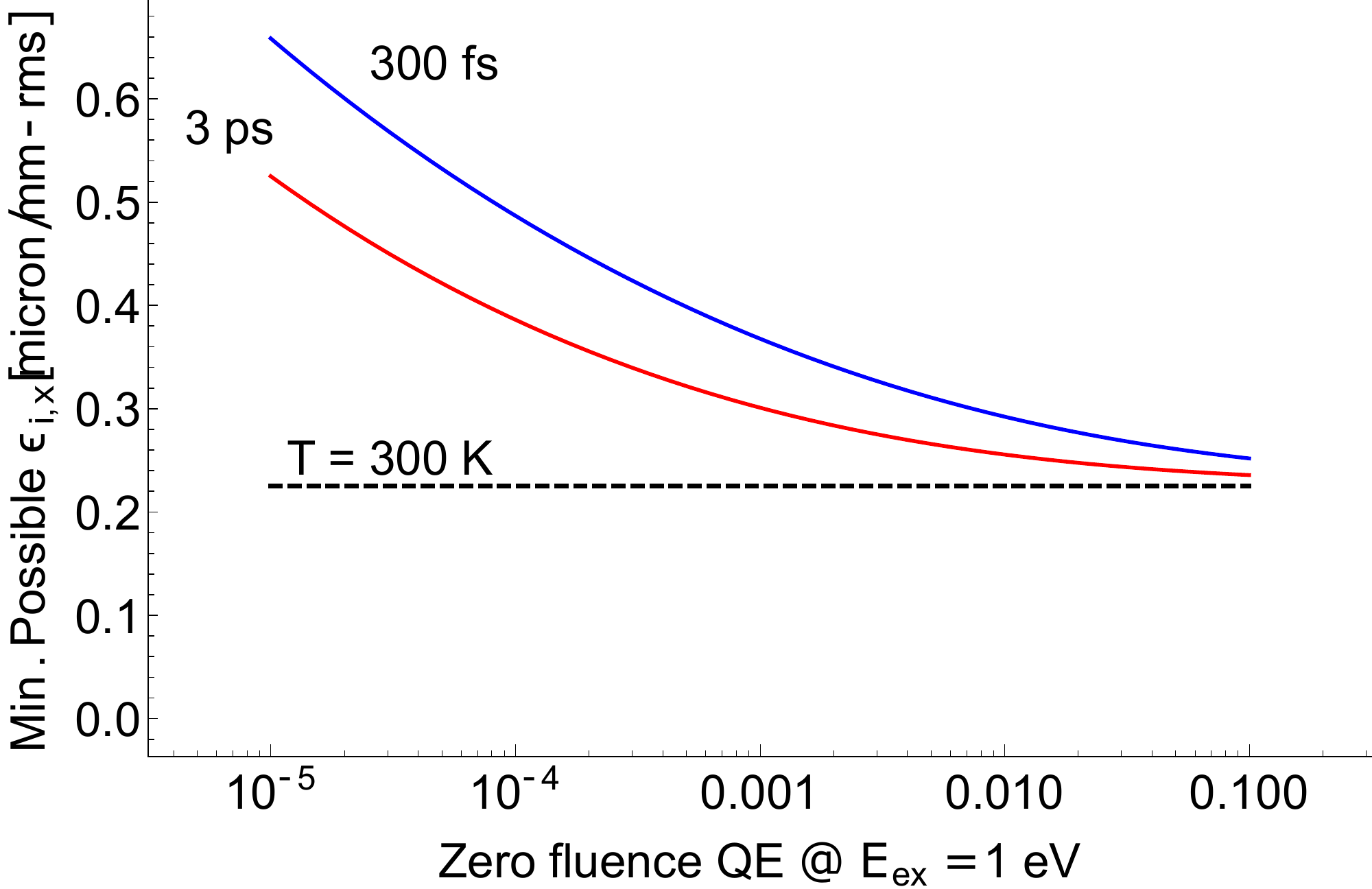}
	\caption{The minimum intrinsic emittance for an extracted charge density of 44 pC/mm$^2$ ($E_0 = 50$ MV/m) as a function of the zero fluence QE at $E_{ex} = 1$ eV.  Other than this QE scale factor, copper material parameters are used. The dotted line represents the emittance for MTE = 26 meV, or the contribution from room temperature. Results shown for $\sigma_t = 3$ and 0.3 ps, with $E_{ex}$ searched from -0.15 to 1 eV.  }\label{minem}
\end{figure} 

\begin{figure}[H]
	\centering
	\includegraphics[width=0.7\textwidth]{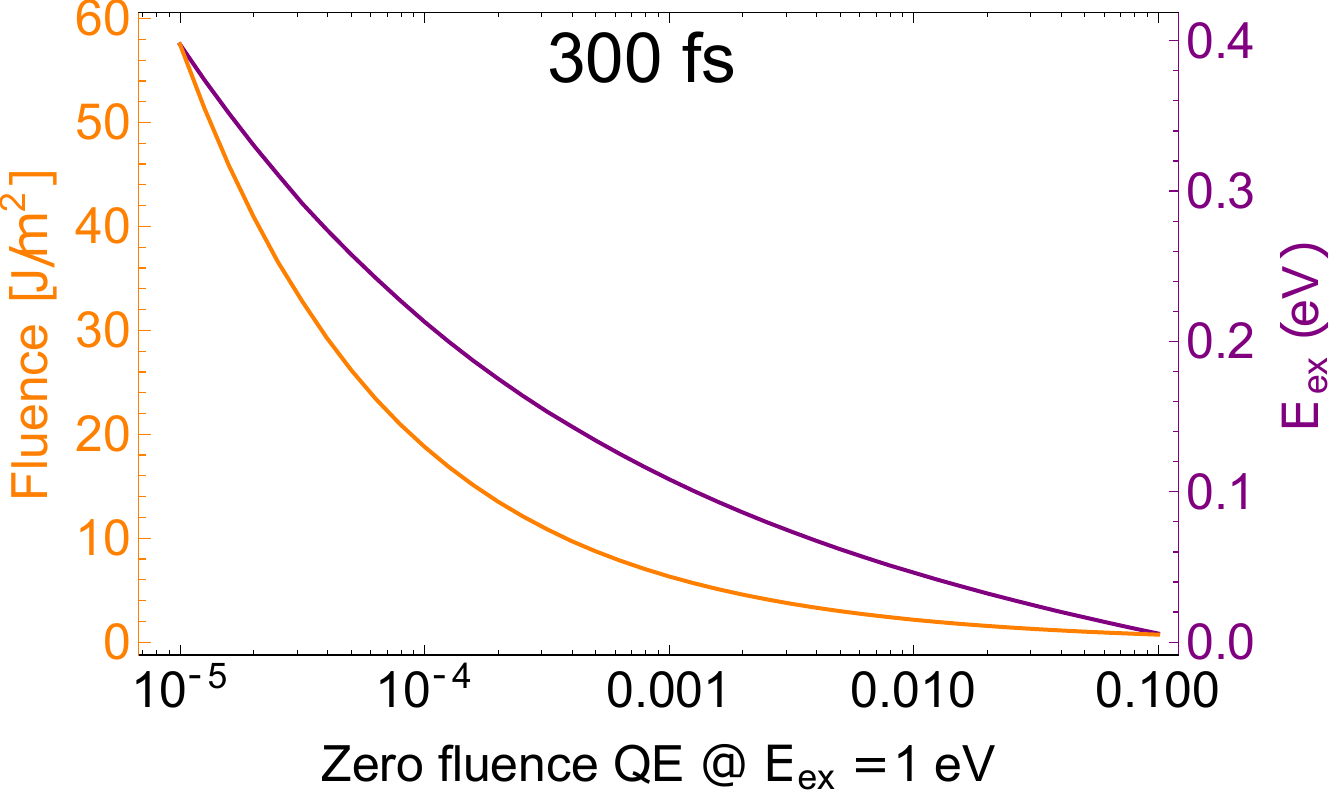}
	\caption{The absorbed fluence and excess energy chosen to minimize the intrinsic emittance for the parameters Fig. \ref{minem} with $\sigma_t = 0.3$ ps.   }\label{fluex}
\end{figure}

With fluence and excess energy constrained by Eq. \ref{required_fluence}, the average MTE,  $\overline{\text{MTE}}\propto \int \text{QE}(t)\text{MTE}(t)S(t, 0) dt $, is no longer a monotonically decreasing function as $E_{ex} \rightarrow 0$. 
Rather, there  will exist a choice of fluence and excess energy that minimizes the average MTE. This minimum MTE will depend strongly on the quantum efficiency scale in Eq. \ref{QEth}. This minimum is depicted in Fig. \ref{fluopt}. Here, a charge density of $\sigma = 0.1 E_0 \epsilon_0 \approx 44 \text{pC/mm}^2$ is extracted with a field of 50 MV/m, for a copper workfunction of $\phi_0 = 4.31$ eV. The reflectivity is held constant at R =0.34, which is approximately true for copper in the wavelength range under consideration. For these material parameters and applied field, the range of $E_{ex}$ from $0 \rightarrow 1$ eV corresponds to $\lambda = 307 \rightarrow 246$ nm. For this case, the minimum MTE is plotted in in Fig. \ref{minem} as a function of the QE scale factor (there shown by setting the QE at F=0 and $E_{ex} = 1$ eV).  To find the optimum fluence and excess energy, they are searched in the range $F \in \left[0, 300\right]$ J/m$^2$ and  $E_{ex}\in \left[-0.15, 1\right]$ eV. It is important to note here that increasing the quantum efficiency scale is equivalent to reducing the required charge density by the same factor, and hence the horizontal scale of Fig. \ref{minem} can be viewed as scanning the parameter $\alpha^{-1}$ for a given QE.

Fig. \ref{fluex} shows the fluence and excess energy chosen in Fig. \ref{QEth} to produce the minimum emittance for the case of $\sigma_t =0.3$ ps. For small QE factors, the optimum excess energy is high, so as to produce sufficient QE to reduce the required fluence and hence the ultrafast heating. For large values of the QE scale factor, the optimum excess energy approaches zero as the influence of ultrafast heating becomes negligible. However, it is noteworthy that the quantum efficiency scale factor must be increased by several orders of magnitude beyond typically achieved values for copper for the effect of ultrafast heating to be fully mitigated.  

\section{Two photon photoemission spectroscopy in an s-band rf gun}

To verify that this effect poses a limit to the intrinsic emittance from cathodes, it is in principle possible to measure the emittance as a function of absorbed fluence using a single pulse to both heat and liberate electrons. However, either using linear (UV) or nonlinear photoemission (NIR), for intensities $>10 \text{ GW/cm}^2$, space charge effects may begin to dilute the emittance. Though these effects may be mitigated via emittance compensation, the emittance must be re-optimized for each fluence setpoint, and the emittance will in general be affected by distortion in the laser spatial mode.

Alternatively, one can employ a variant of time-resolved two photon photoemission spectroscopy (TPPS) to measure this effect using both a pump and probe pulse. In short, in this proposed scheme, a pump laser pulse of high intensity and long wavelength drives the photocathode heating, while a probe laser of multiple ps duration photoemits a beam with a time dependent thermal emittance, smaller at later times, due to the electron phonon coupling. We envision this experiment to take place in a standard 1.6 cell normal conducting RF gun with a downstream slice transverse emittance diagnostic. 

In this proposed experiment, we set the pump pulse intensity to be $ 10$ mJ/cm$^2$ (absorbed), $\sigma_t=300$ fs, as in the calculations above, corresponding to 13 GW/cm$^2$ peak intensity. The wavelength is set 800 nm. Though the reflectivity of Cu is poor at 800 nm, the cathode may be AR coated without effect on the intrinsic emittance \cite{MusumeciCultreraFerrarioEtAl2010}. For a uniform pump beam of radius 100 micron at the photocathode, this requires only $\sim 3$ $\mu$J of absorbed energy. 

The probe pulse is a low intensity UV pulse with multiple ps duration with variable delay relative to the probe pulse. The probe wavelength is set to yield $E_{ex} = 0.36$ eV (green curves in Fig. \ref{etherm}), which provides a balance between sufficient QE and small enough MTE to resolve the induced temperature modulations. For a probe pulse of the same radius as the pump, a charge of 0.1 pC, and an assumed linear QE of $10^{-6}$, this requires a probe fluence of $\sim 1$ mJ/cm$^2$ at the cathode, which with a 3.5 ps pulse length yields a maximum induced temperature of only $< 600$ K, which is small compared to the temperature induced by the pump, which is rougly 3000 K at the peak. 

   Considering that we seek to measure a modulation of the beam temperature in the probe electron beam, we must determine whether the modulation persists downstream at the emittance diagnostic, including any effects of space charge and temperature diffusion within the probe electron beam. To do this, we employ the space charge tracker General Particle Tracer \cite{gpt} to perform start-to-end simulations of the process. In this case, the temporal distribution is set to a flat top with length 3.5 ps, which is approximately the time required for the peak electronic temperature to reach a minimum. A linear ramp in the MTE from the cathode, corresponding approximately to the emittance profile ramp given in Fig. \ref{etherm} (MTE from $300 \rightarrow 127$ meV), using a low-discrepancy Hammersley sequence to minimize effects of shot noise with $10^5$ macroparticles. Only smooth space charge forces are considered, which is valid considering the low density and high temperature of the probe beam. Any pump-driven photoelectron beam is not modeled, and is foreseen to be blocked by using a mask downstream of the slice-emittance diagnostic deflection cavity.      

\begin{figure}[h]
	\centering
	\includegraphics[width=0.7\textwidth]{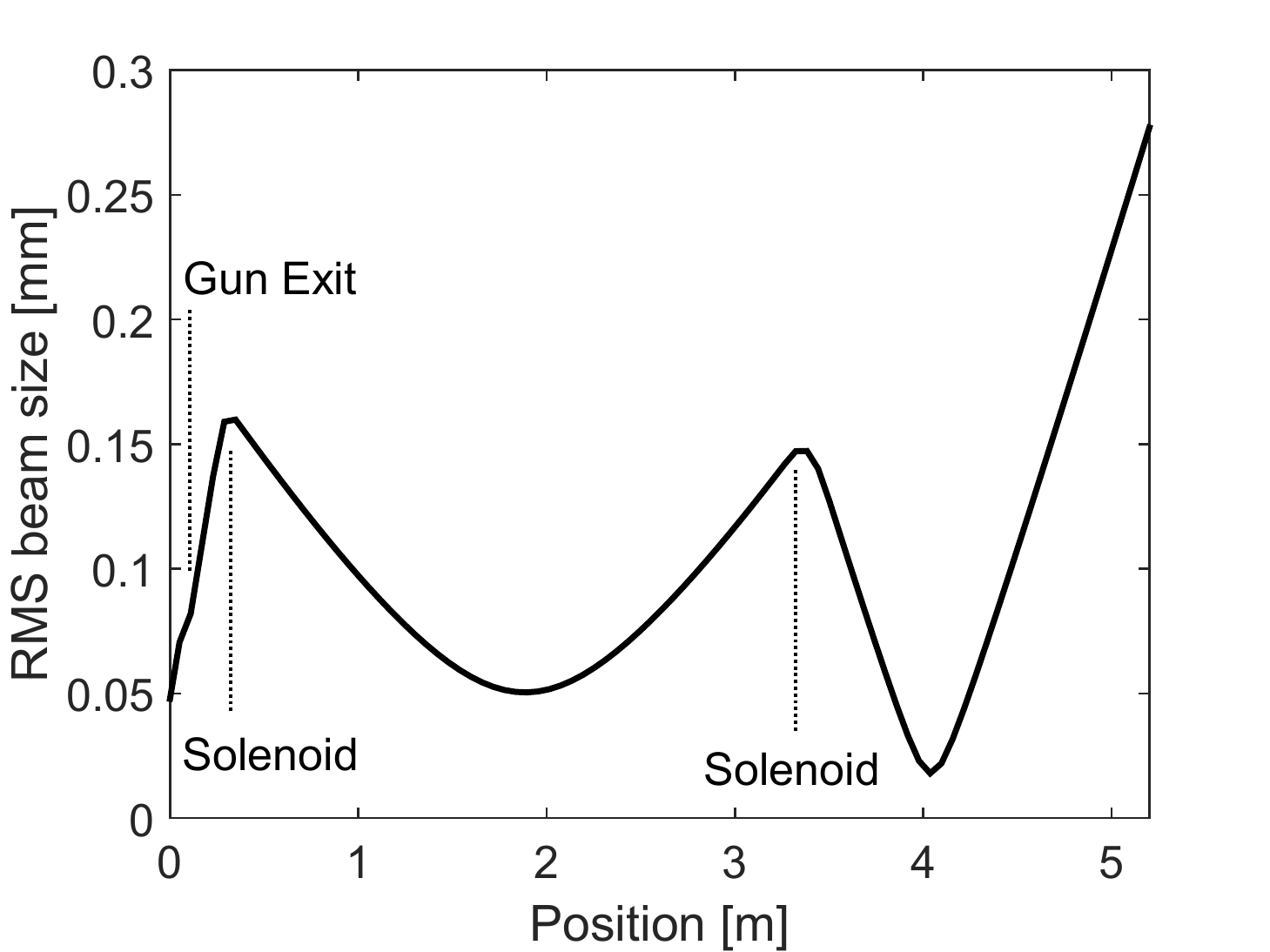}
	\caption{Beam transverse size as a function of position along the UCLA Pegasus beamline in the TPPS measurement scheme. }\label{spotsize}
\end{figure}

We use the UCLA Pegasus 1.6 cell gun and beamline (without linac) as the layout for the experiment. The beam energy is 4.1 MeV, and the measurement point is set at 5.2 m downstream of the cathode. We foresee the use of the TEM-grid inverse-pepperpot emittance technique \cite{LiRobertsScobyEtAl2012} with a slit and deflection cavity for the use in the experiment, which requires a tight beam focus upstream of the TEM grid for point-projection imaging. The beam size along the beamline in this configuration is shown in Fig. \ref{spotsize}.

\begin{figure}[h]
	\centering
	\includegraphics[width=0.7\textwidth]{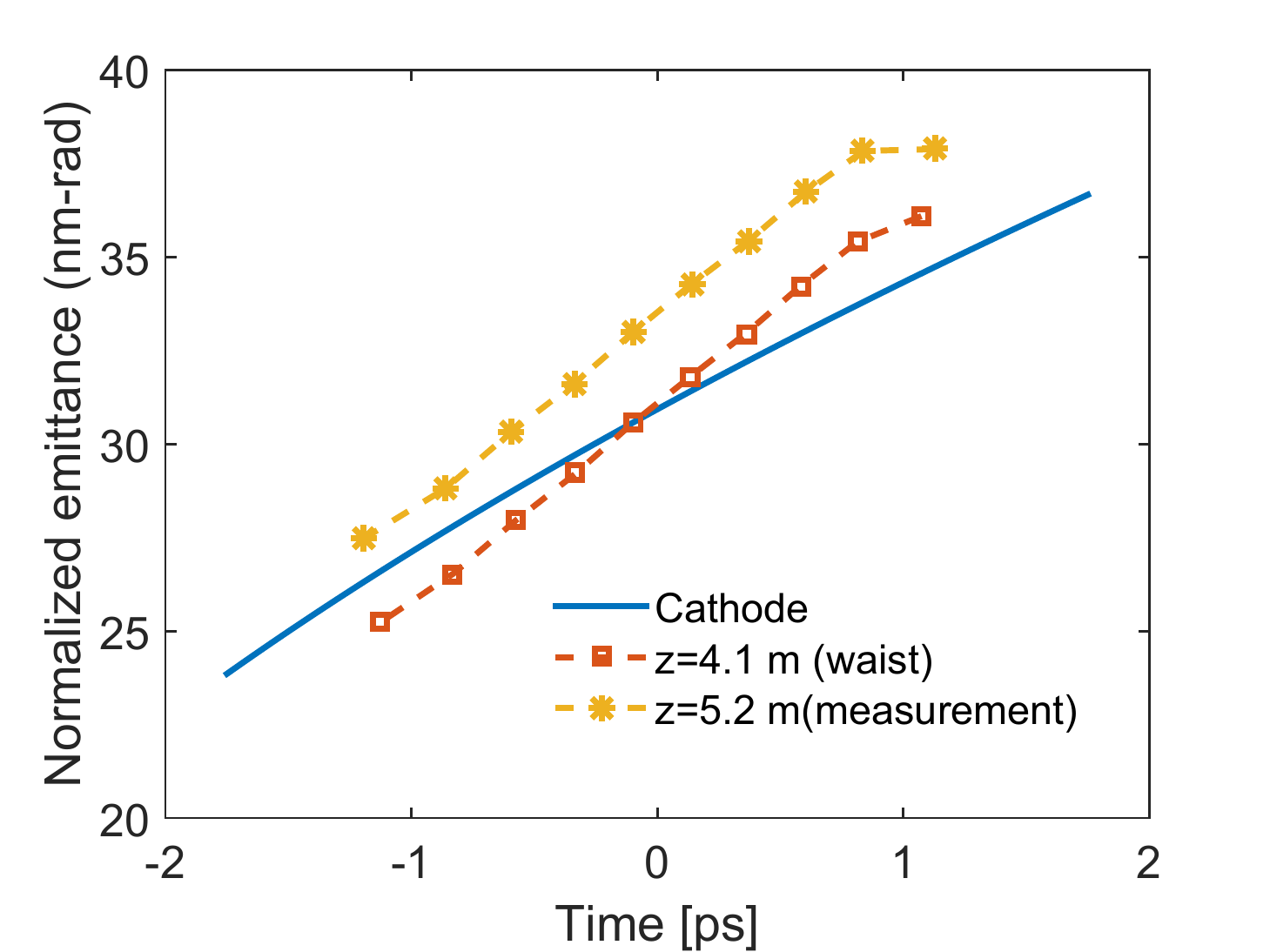}
	\caption{Slice emittance of the beam vs intrabunch temporal coordinate at cathode, just prior to the point-projection waist, and at the measurement position. }\label{evst}
\end{figure}

The emittance as a function of the temporal coordinate of the bunch at the cathode, just prior to the waist, and at the measurement point is shown in Fig. \ref{evst}. Note that both the bunch length and emittance modulation are largely preserved up to the waist. After the waist the beam experiences a global emittance increase, but the emittance slope remains roughly constant. Thus, the temperature modulation of the MTE induced from ultrafast laser heating can persist and would be measurable with a slice emittance diagnostic.

\section{Conclusions}

 In this work, we have shown that the intrinsic emittance of metallic cathodes driven with ps-scale and shorter laser pulses can be directly limited by the ultrafast heating of the electronic distribution. We applied the widely successful two temperature model to calculate this heating for multiple cases that might be typical of RF photoinjector operation. In doing so, we find the full mitigation of this heating effect must come from a reduction of the required laser fluence via QE increase. Partial mitigation can be found via the use of longer, multiple-ps duration laser pulses, where applicable, as such pulses make use of the ps electron-phonon equilibration timescale. Finally, we propose an experiment to measure the laser induced time dependent modulation of the cathode intrinsic emittance using a variation of two-photon time resolved photoelectron spectroscopy but using standard photoinjector diagnostics downstream of a 1.6 cell RF photogun.

\section{Acknowledgments}  
Helpful conversations with Renkai Li and Daniele Filippetto are gratefully acknowledged. This work was supported by the DOE STTR award DE-SC0009656 and NSF PHY-1415583.

\end{document}